\def\N{\mathbb{N}}
\def\td{\mbox{d}}		
\def\Pr{\mbox{Pr}}	

\documentclass[10pt,a4paper]{article}
\usepackage{amsmath}
\usepackage{amsfonts}
\usepackage{epsfig}
\def\appendix{\par
    \setcounter{section}{0}\setcounter{subsection}{0}
    \def\thesection{\Alph{section}} \section*{Appendix}
}
\begin{document}

\title{Properties of a non-equilibrium heat bath}
\author{Aditi Simha, R. M. L. Evans and A. Baule \\ \\
{\it School of Physics and Astronomy, University of Leeds, LS2 9JT, U.K.}}
\date{20th September 2007}

\maketitle

\begin{abstract}
At equilibrium, a fluid element, within a larger heat bath, receives random impulses from the bath. Those impulses, which induce stochastic transitions in the system (the fluid element), respect the principle of detailed balance, because the bath is also at equilibrium. Under continuous shear, the fluid element adopts a non-equilibrium steady state.
Because the surrounding bath of fluid under shear is also in a non-equilibrium steady state, the system receives stochastic impulses with a non-equilibrium distribution. Those impulses no longer respect detailed balance, but are nevertheless constrained by rules. The rules in question, which are applicable to a wide sub-class of driven steady states, were recently derived [R. M. L. Evans, Phys.\ Rev.\ Lett.\ {\bf 92}, 150601 (2004); J. Phys.\ A: Math.\ Gen.\ {\bf 38}, 293 (2005)] using information-theoretic arguments. In the present paper, we provide a more fundamental derivation, based on the uncontroversial, non-Bayesian interpretation of probabilities as simple ratios of countable quantities. We apply the results to some simple models of interacting particles, to investigate the nature of forces that are mediated by a non-equilibrium noise-source such as a fluid under shear.
\end{abstract}



\section{Introduction}
\label{intro}

Throughout the literature on non-equilibrium statistical mechanics, while countless models of non-equilibrium systems have been defined, the stochastic noise that impinges on those systems is, with very few exceptions \cite{Hayashi07}, assumed to emanate from an {\em equilibrium} heat bath. In reality, a fluid element, that lies in the middle of some driven complex fluid under continuous shear, is buffeted by the surrounding fluid elements, and thus receives random impulses with a distinctly non-equilibrium distribution. In this paper, we investigate the statistical properties of the noise that emanates from such a non-equilibrium reservoir.

Non-equilibrium steady states of matter are ubiquitous, and as varied as equilibrium states. For instance, when continuously sheared at a high rate, a concentrated solution of phospholipids chooses to arrange its molecules, onion-like, into concentric spherical layers \cite{Roux}. Why one such macrostate is statistically favoured over another is a matter of speculation, since the relevant laws of statistical mechanics have not been rigorously derived away from equilibrium. 

Over the last couple of decades, much theoretical effort has been directed towards finding a non-equilibrium counterpart to Boltzmann's law, to provide a general formula for the occupancy of microstates. Despite some promising theoretical approaches based on Jaynes's information-theoretic derivation of statistical mechanics \cite{Jaynes57,JaynesBook}, no such result has been forthcoming without some approximation: typically a coarse-graining step, or a near-equilibrium assumption. This does not imply that exact and general results are unobtainable in a non-equilibrium context, as witnessed by Jarzynski's non-equilibrium work theorem \cite{Jarzynski}, but only that the particular problem of microstate occupancy has proven difficult to tackle.

Recently, it was noticed that an exact result could be obtained \cite{PRL,JPhysA} if one asks, not for the probability of occupying a particular microstate, but for the probability of moving between a particular pair of microstates, i.e.~the transition rate. Traditionally, rather than being {\em derived}, transition rates have been {\em defined} as the starting point for modelling non-equilibrium processes. By contrast, {\em equilibrium} statistical mechanics begins from a definition of an ensemble, and has much to say about the transition rates that are allowed in a model; through the {\em derived} principle of detailed balance. In the recent articles \cite{PRL,JPhysA}, a non-equilibrium counterpart to the principle of detailed balance (NCDB) was derived. The starting point was Jaynes's information-theoretic principle of maximum-entropy inference (MaxEnt) \cite{Jaynes57,JaynesBook}, which is a controversial topic in its application to non-equilibrium systems, since non-equilibrium ensembles of phase-space trajectories are often ill-defined (a counter-example being \cite{Bruers07}). It is therefore generally unclear under what circumstances MaxEnt can be regarded as exact. By contrast, the derivation \cite{PRL,JPhysA}, {\em from} MaxEnt, of non-equilibrium transition rates, is robust, since no approximate steps (such as the usual coarse-graining approximations) were used. In other words, if MaxEnt is correct in the context of non-equilibrium steady states, then so is NCDB \cite{PRL,JPhysA}. The MaxEnt recipe asserts that the {\em path entropy}
\begin{equation}
\label{MaxEnt}
	S = - \sum_\Gamma p(\Gamma) \, \ln p(\Gamma)
\end{equation}
of a distribution $p(\Gamma)$ of phase-space paths $\Gamma$ should be maximised, subject to certain constraints. In the present article, we shall not repeat the derivation of NCDB from Eq.~(\ref{MaxEnt}), which can be found in full in Ref.~\cite{JPhysA}, but shall present a more intuitive derivation of Eq.~(\ref{MaxEnt}) itself (the MaxEnt recipe) for sheared steady-states. Our derivation will include some precise definitions that are usually absent from non-equilibrium analyses. In particular, we shall discuss the meaning of a non-equilibrium ensemble and the properties of a non-equilibrium heat-bath, since the NCDB quantifies the coloured noise that emanates from a heat bath which is itself under continuous shear (or otherwise mechanically driven).

Before proceeding further, we should address the question of whether a steady state can actually exist in a fluid under continuous shear, since only then do we have a real physical system to which our exact derivation can be applied. Certainly one can invent physically motivated models of complex fluids for which such states exist, involving stochastic equations of motion with dissipative forces \cite{Blythe07} as well as conservative and random ones. An isolated fluid of Newtonian particles, on the other hand, must gradually heat up, since work is done by the driving force. So the system's state is not truly steady, but only quasi-steady. An experimentalist observing such a system would measure transition rates by counting the occurrences of a given transition over a time interval during which the temperature remained relatively constant. Sometime later, when the system has heated up, the measurements could be repeated at a new, higher temperature. If the temperature increases so rapidly that it is non-constant on the timescale of the transitions themselves, the observer would be unable to report any steady-state data, and our theory would also be inapplicable to that system. Alternatively, the experimenter might enforce a truly steady state by having the system open to the laboratory, or enclosed between temperature-controlled rheometer plates. The troublesome fluid, for which no steady  state could be measured adiabatically, will now exhibit steep temperature gradients on the spatial scale of the microscopic structure, so that the experimentally-determined rates become strongly dependent on the geometry of the boundaries and, again, no general fluid properties are obtainable. 

In practice, experimentalists do observe non-equilibrium steady states of many complex fluids, and are able to quote bulk properties that are insensitive to boundary conditions and measuring times. For instance, in many complex fluids, a solvent acts as a thermostat for the much larger solute particles that exhibit the interesting non-equilibrium behaviour. The space- and time-dependence of the solvent's temperature can always be neglected in such cases, since the work done does not drive the solvent significantly far from equilibrium, although the complex fluid as a whole may be in a highly non-equilibrium state. It is only for boundary-driven systems in which the microscopic dynamics are not significantly affected by the rate of heating (in the adiabatic case), or by the temperature gradients (in the thermostatted case) that one can meaningfully quote such bulk properties. Those are the systems for which we shall construct our theory. The condition for such an assumption to be good is elucidated by considering a counter-example: granular media. In flowing granular media, heat is dissipated in every grain, so that collisions are inelastic; it is crucial to this system's behaviour that it is irreversible at the level of the grain size (the scale of interest) even though the grains are each composed of many Newtonian particles performing reversible dynamics. By contrast, the temperature-dependent dynamics of the amphiphiles in a sheared onion phase is not sensitive to the (small) {\em rate of change} of temperature.
Hence, for the large class of complex fluids that experimentally exhibit far-from-equilibrium steady states without significant heating, we may safely neglect large-scale temperature gradients whilst simultaneously assuming microscopic reversibility (Newtonian laws of motion), without breaking time-translation invariance, even though consistency demands that such systems should, in truth, very slowly heat up. Having made that assumption in defining our ensemble, no approximations are thereafter required to derive the exact consequences of the ensemble's dynamics, in terms of the detailed-balance-like constraints on transition rates. 

The new derivation of Eq.~(\ref{MaxEnt}) is presented in section \ref{foundations} following a discussion, in section \ref{prelim}, of the equivalent concepts at equilibrium. We shall then proceed to investigate the implications for the effective forces between interacting particles that are driven by a non-equilibrium heat bath. That investigation begins, in section \ref{exclusion}, with a simple model of exclusive site-hopping, and is discussed in terms of more general interactions in section \ref{Newton}, where we show that the spatial and temporal symmetries built into the formalism imply that the heat-bath-mediated effective forces respect Newton's third law of motion.

\section{Preliminaries: A system at equilibrium}
\label{prelim}

We begin by discussing familiar concepts of equilibrium statistical mechanics, in a way that can be straightforwardly generalised in the next section.

Consider a system (a fluid) composed of a set of interacting particles that obey microscopically reversible dynamics, and are sufficiently mobile (in a phase-space sense) to explore a representative set of the states available to them (i.e.~the system is ergodic). Let this system be weakly coupled to a large number of similar systems (an ensemble). We can imagine the ensemble to be a single huge super-system, that is divided by imaginary planes into a grid-like arrangement of systems. By ``weakly coupled" we mean that each system is much larger than the largest correlation length $l_{\rm corr}$, so that neighbouring systems are uncorrelated. Thus the only significant interaction between different systems is via conserved quantities such as energy, since a change in one system's share of such  quantities necessitates an opposite change in the other systems' share, no matter how far they are separated.

We see that each system is subject to a source of noise coming from the ``reservoir" or ``heat bath" constituted by the rest of the ensemble. Even if the equations of motion governing the entire super-system (the ensemble) are deterministic, an individual system, in the presence of the ``heat bath" (the rest of the ensemble), is described by dynamics with a stochastic element. We ask: what is the likelihood that a particular system, currently in state $a$, receives a stochastic impulse of the appropriate size and direction such that it is transformed into state $b$? In other words, what are the transition rates $\omega_{ab}$ that describe this stochastic dynamics? We cannot answer this question in full without an exact specification of the ``microscopically reversible dynamics" that governs the system and reservoir. However, we do know, in general, that the transition rates are subject to a number of constraints due to the statistics of the reservoir. In particular, they are constrained to obey detailed balance:
\begin{equation}
  \omega_{ab}/\omega_{ba}=e^{-(E_a-E_b)/k_BT}
\end{equation}
where $E_i$ is the energy of microstate $i$. The fact that the statistics is subject to this set of constraints arises only from the following properties of the system and reservoir: (i) ergodicity, (ii) microscopic reversibility (iii) they are in a statistically steady state (iv) energy is a conserved quantity that can be exchanged between systems.

Now let us apply some constant rate of shear to the super-system, by doing work to impart constant relative motion to a pair of parallel plates located at the super-system's distant boundaries. Thus, we are imposing a constraint on the amount of shear-flux experienced by the ensemble as a whole, but have in no way modified the equations of motion governing the individual systems. Each system still has the same Hamiltonian as in the equilibrium case, and respects the same laws of physics. As a result, the same four crucial properties listed above still apply, with the addition of an extra conserved quantity that can be partitioned between systems: the net shear rate. The constraints on equilibrium transition rates, embodied by the principle of detailed balance, followed as a consequence of those four crucial properties. Hence, the same degree of constraint must also prevail in the weakly-coupled boundary-driven ensemble; the statistics of the reservoir should again yield detailed-balance-like relations, now modified by the shear rate. Those were the relations calculated in Refs.~\cite{PRL,JPhysA,PhysicaA}.

\section{The foundations of the derivation}
\label{foundations}

The derivation of the non-equilibrium counterpart to detailed balance (NCDB) (set out in greatest detail in Ref.~\cite{JPhysA}) relied on Jaynes's \cite{Jaynes57,JaynesBook} information-theoretic interpretation of non-equilibrium statistical mechanics, using Shannon's \cite{Shannon} information entropy for the set of phase-space trajectories. This is an increasingly popular basis for non-equilibrium calculations \cite{Ghosh06}, but remains controversial in the non-equilibrium context, for a number of reasons: 
{\bf (i)} Due to Jaynes's discussions \cite{Jaynes57,JaynesBook} the concept of information entropy is strongly linked to the Bayesian interpretation of probability: that probability is a subjective quantity \cite{Dewar05}, describing the observer's state of ignorance, rather than countable realisations of the physics. 
{\bf (ii)} The meaning of the ``ensemble" in much of the literature is unclear; we have no concrete interpretation for what prior set of trajectories is being considered --- whether the set includes unphysical trajectories, to be removed {\it a posteriori} --- stochastic or Hamiltonian trajectories, etc.
{\bf (iii)} The application of this kind of formalism often involves all kinds of implicit coarse graining and mean field approximations that are not easy to pinpoint in the calculations.
In fact, none of these features is necessary to calculations on non-equilibrium ensembles, as can be seen below, where a concrete interpretation of a driven steady state ensemble is set out, providing a more rigorous basis for the axioms used in Refs.~\cite{PRL,JPhysA,PhysicaA}.

\begin{figure}
  \epsfxsize=6cm
  \begin{center}
  \leavevmode\epsffile{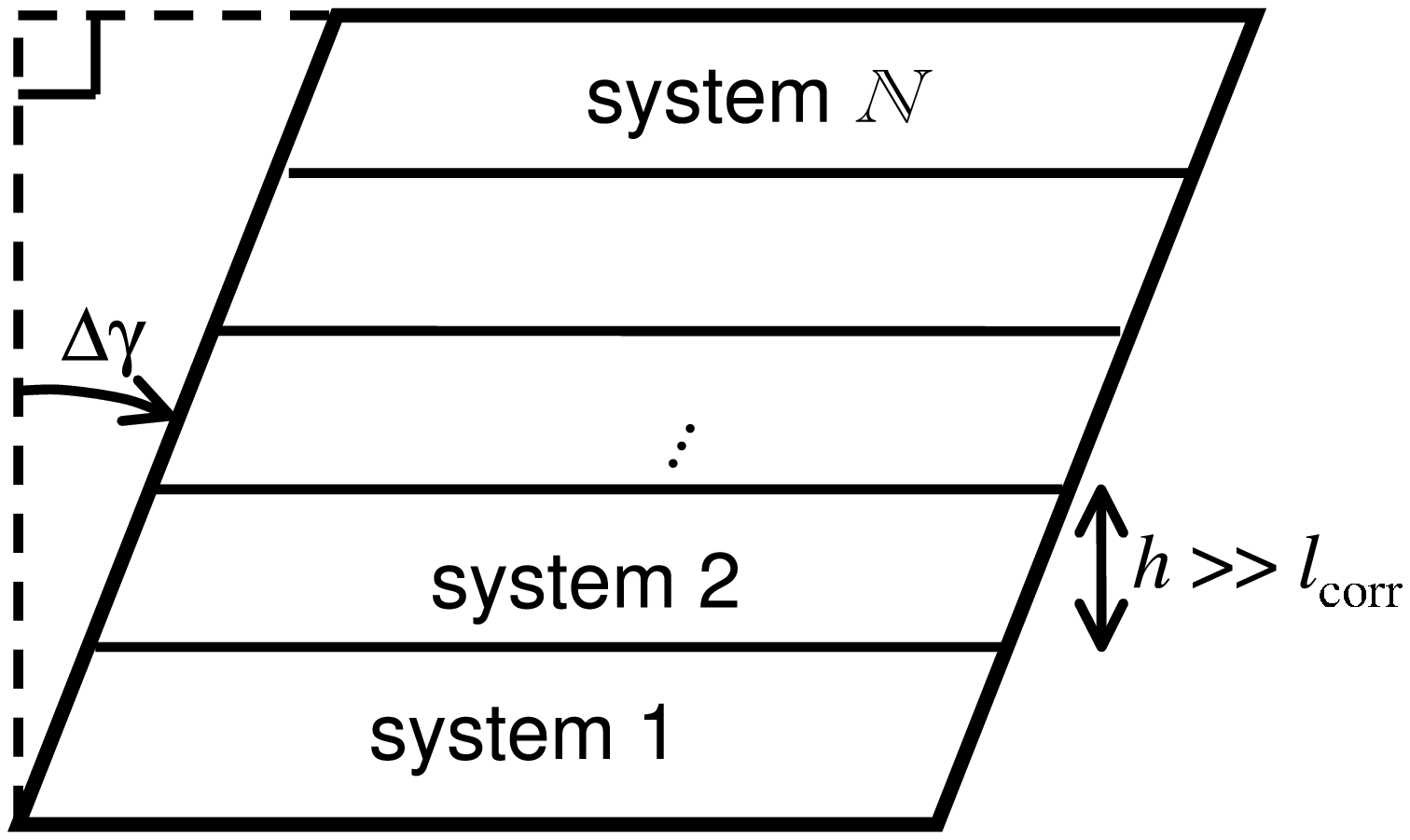}
  \caption{\label{ensemble}The ensemble of systems, stacked up in the shear-gradient 
	direction, to form a super-system under shear.}
\end{center}
\end{figure}

As before, consider a very large quantity of fluid under shear: a ``super-system" which we imagine to be divided into $\N$ sub-regions or ``systems". So many systems comprise the super-system that it constitutes an ensemble. The ensemble is depicted in Fig.~\ref{ensemble}, which shows that the systems are stacked up in the velocity gradient direction, and that each system is much larger than the largest correlation length $l_{\rm corr}$, so that a negligible fraction of each system is correlated with the states of neighbouring systems.

In time $\tau$, this super-system undergoes a total amount of shear $\langle\gamma\rangle$. We take $\tau$ to be much longer than any transients or correlation times of the system, so that the time-span under consideration is dominated by steady-state behaviour. The shear experienced by the $i$th system in this time is $\gamma_i$, so that 
\begin{equation}
\label{gamma}
	\sum_{i=1}^{\N}\gamma_i = \N \langle\gamma\rangle,
\end{equation}
and we see that the time- and ensemble-averaged shear rate is $\left\langle\overline{\dot{\gamma}}\,\right\rangle=\langle\gamma\rangle/\tau$. Furthermore, each of the systems has an identical Hamiltonian. 

What is the probability $p(\Gamma)$ that, during time $\tau$, a particular system follows a microscopic phase-space trajectory $\Gamma$? The trajectory $\Gamma$ represents the system's entire history of behaviour, down to the microscopic level, during the duration $\tau$. Recall that, during this time, the system is subject to noise from its non-equilibrium surroundings, so that $\Gamma$ is not fully determined by its initial state. 

To answer the question and find the distribution of trajectories $p(\Gamma)$, we do not need to appeal to Jaynes's Bayesian interpretation of probabilities, nor do we need to coarse-grain the system and find dynamical equations for locally averaged quantities. Instead, we can follow Gibbs's exact calculation for the statistical weight of an entire ensemble as follows \cite{Mandl}.

Assume that the ensemble is so large that each possible trajectory $\Gamma$ is realised in many of the systems, i.e.~many systems experience identical histories. Then the number $n_\Gamma$ of systems that follow trajectory $\Gamma$ is $n_\Gamma = \N\, p(\Gamma)$. The statistical weight of the entire ensemble is the number of ways of permuting these differently-experienced systems, which is
\[
	\Omega_{\N} = \frac{\N !}{\prod_\Gamma n_\Gamma !}.
\]
If we now make many copies of the entire ensemble, the set of ensembles will be dominated by those with the largest statistical weight. This is maximised by maximising its logarithm, the ``path entropy" of the ensemble,
\[
	S_\N \equiv \ln \Omega_\N = -\N \sum_\Gamma p(\Gamma) \, \ln\, p(\Gamma),
\]
so the path entropy per system is
\setcounter{equation}{0}
\begin{equation}
\label{entropy}
	S = - \sum_\Gamma p(\Gamma) \, \ln\, p(\Gamma).
\end{equation}
\setcounter{equation}{3}
We recognise the Shannon entropy \cite{Shannon}, and realise that this is the quantity maximised by the ensemble (subject to physical constraints) simply as a result of combinatorics. Here, we have reproduced Gibbs's argument, which applies to any probability distribution. If we are considering a set of physical systems, then the maximisation of $S$ with respect to $p(\Gamma)$ must respect the laws of physics, i.e.~the paths $\Gamma$ are {\it a priori} assigned zero weight for unphysical paths. 

We have been compelled to consider high-dimensional phase-space trajectories, rather than microstates or configurations of the system because the laws of physics give rise to temporal correlations, i.e.~the dynamical evolution of the system at one time is coupled to its evolution at another time. The fact that the order of events can usually be disregarded in statistical mechanics is a special property of the equilibrium ensemble. It is partly a result of the statistical de-coupling of the momentum degrees of freedom in canonical equilibrium, which is a happy accident that cannot be assumed in general (e.g.~for sheared fluids).

Having derived the path entropy, we note that it can be maximised (subject to constraints) to obtain the distribution of paths followed by systems in a weakly-coupled boundary-driven steady state, or can equally well be applied to the trajectories in equilibrium systems.
The space of all possible trajectories is a high-dimensional continuum, so that any discussion of probability distributions, defined on that space, requires us to consider an appropriate {\em measure}. The simpler problem in equilibrium statistical mechanics, to define an appropriate measure on the space of {\em microstates}, is solved by the principle of equal {\em a priori} probabilities, implying a uniform measure on Hamiltonian phase space, a principle that derives from Liouville's theorem. The appropriate measure on the higher-dimensional space of all trajectories is more difficult to quantify. Happily, however there is no need to do so for the present derivation. Whatever is the (unknown) correct measure for the set of trajectories of a dynamical system {\em at equilibrium}, that remains the correct measure for systems in the driven ensemble, since the equations of motion governing the systems are {\em the same} in the driven and equilibrium ensembles. This is an important consequence of the way in which this particular class of steady states is driven (mechanically driven, at the boundary, by weak coupling to other systems). Thus, the trajectories in the driven ensemble have the same measure as in the flux-free equilibrium ensemble, and are only {\em re-weighted} by the posterior flux constraint (Eq.~(\ref{gamma})), yielding an extra factor $e^{\nu\,\gamma(\Gamma)}$ after maximisation of Eq.~(\ref{entropy}), i.e.
\begin{equation}
\label{proportional}
	p^{\rm driven}(\Gamma) \propto p^{\rm equilib}(\Gamma)\: 
e^{\nu\,\gamma(\Gamma)}
\end{equation}
with Lagrange multiplier $\nu$ chosen to fix $\langle\gamma\rangle$ (where all quantities are still implicitly $\tau$-dependent at this point). The resulting probability that a trajectory will contain a particular transition $a\to b$ then follows from the derivation in Refs.~\cite{PRL,JPhysA}, yielding a relation between the transition rate $\omega_{ab}$ in the driven ensemble and that at equilibrium. Hence, there is a one-to-one mapping between a system's transition rates in the equilibrium and driven ensembles. Since the equilibrium rates cannot be freely chosen, but are constrained to respect the principle of detailed balance, the one-to-one mapping implies exactly the same degree of constraint on the choice of non-equilibrium rates.

Specifically, the mapping  between the  transition rate $\omega_{ab}^{\rm dr}$ from state $a$ to state $b$ and the rate $\omega_{ab}^{\rm eq}$ at equilibrium (derived in \cite{JPhysA} and alternatively from Eq.~(\ref{proportional}) in the Appendix), can be written as
\begin{equation}
  \label{can1}
  \omega_{ab}^{\rm dr}=\omega_{ab}^{\rm eq}\, \exp \left[\nu J_{ab} \Delta t
   +q_{b}(\nu) -q_{a}(\nu) -Q(\nu)\Delta t \right]
\end{equation}
where $J_{ab} \Delta t$ is the net increment of flux gained by the transition, $\Delta t$ is the discrete time step (which vanishes for the continuous-time dynamics considered below), and $Q(\nu)$ is a property of the system's steady state, akin to a thermodynamic potential, and is related to the average current by $dQ/d\nu =J$. (Note that all quantities in Eq.~\ref{can1} are independent of the arbitrarily long duration $\tau$.) Finally, $q_{a}$ is a property of the microstate $a$, defined as 
\begin{equation}
  \label{can2}
  q_{a}(\nu)=\lim_{\tau\rightarrow \infty} \left[ \ln\int 
  p_{\tau}^{\rm   eq}(J|a)\,e^{\tau \nu J} dJ -\tau Q(\nu) \right]
\end{equation}
and measures the system's propensity to exhibit flux in future, given the current microstate. Here $p_{\tau}^{\rm eq}(J|a )$ is the probability that the system, currently in state $a$, subsequently acquires net flux $J$ in time $\tau$ in an equilibrium bath.

If $J$ denotes a {\em shear} flux, we can identify $J \tau =\gamma$. However, in the following section, we apply the above formula to the simplest model of an interacting particle system: two-particle exclusion, in which the flux in question is not shear but mass transport. In this case, the weak coupling criterion for a boundary-driven ensemble becomes more difficult to envisage than for the realistic case of shear flow considered above. It should therefore be remembered that the model is an abstract application of the formalism, embodying some of the important features of more complex interacting systems.

\section{Two-particle exclusion model}
\label{exclusion}

Consider two particles on a one-dimensional lattice, each capable of hopping one site to the right or left, provided
that site is vacant, with rates $\omega$ that are equal at equilibrium, as a result of symmetry and detailed balance. When this system is weakly coupled to an ensemble of the boundary-driven type discussed in section \ref{foundations}, with particle flux replacing shear flux, the rates are given by eqns.~(\ref{can1},\ref{can2}).
 
\begin{figure}
  \begin{center}
  \epsfig{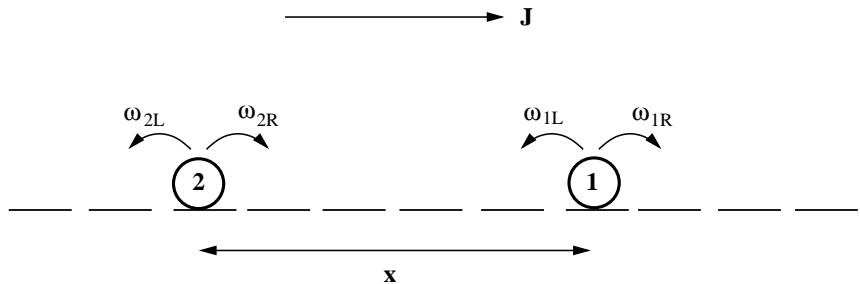}
  \caption{\label{sep}The two-particle exclusion model.}
\end{center}
\end{figure}

The state of the system at any given time is defined by the separation $X$ between the particles (see Fig.~\ref{sep}), and by their centre-of-mass position which, on average, drifts to the right in the driven ensemble but, due to the system's translational symmetry, cannot affect the hopping rates. Since the average flux is some function of the parameter $\nu$, the transition rates can be written in terms of this parameter, for any microstate of the system labelled only by the particle separation $X$. We shall see that $\nu$ plays the role of an external stress, driving the ensemble.

Consider the particles initially at a separation $X$ ($> 1$). Since either particle can hop right or left there are, in total, four allowed transitions, leading to microstates with separation $X-1$ or $X+1$. These include
hops to the right and left for particle $1$ while particle $2$ remains static (labelled $1R$ and $1L$ respectively) and similarly those for particle $2$ with $1$ fixed (2R and 2L ). Simultaneous hopping of both particles is not considered as it occurs with vanishing probability in continuous-time dynamics.
Transitions involving a hop to the right carry a net flux 
$J_{ab}\Delta t=1/2$ (since the centre of mass is displaced by half a unit in such transitions). Similarly for hops to the left $J_{ab}\Delta t=-1/2$. In continuous time, $\Delta t \rightarrow 0$ and eqn.(\ref{can1}) yields
\begin{eqnarray}
  \omega_{1R}&=&\omega\,e^{\nu/2}e^{q_{X+1}(\nu)-q_{X}(\nu)}
\label{omega1R} \\
  \omega_{1L}&=&\omega\,e^{-\nu/2}e^{q_{X-1}(\nu)-q_{X}(\nu)}\\
  \omega_{2R}&=&\omega\,e^{\nu/2}e^{q_{X-1}(\nu)-q_{X}(\nu)}\\
  \omega_{2L}&=&\omega\,e^{-\nu/2}e^{q_{X+1}(\nu)-q_{X}(\nu)}
\label{omega2L}
\end{eqnarray}
We now require the function $q_X(\nu)$ which, from Eq.~(\ref{can2}), depends on the {\em equilibrium} dynamics (via $p_{\tau}^{\rm   eq}(J|a)$).

From the initial state at time $t=0$, the probability that a transition, into one of the four allowed states $\alpha$, occurs within the time interval \mbox{$t\to t+\td t$}, is given by \mbox{$p_{\alpha}(t)\,\td t = \omega\, e^{-4\omega t}\,\td t$} (since each of those transitions has a rate $\omega$). Following the transition, the system's evolution is governed by the propagator of its final state $\alpha$. Therefore, for an initial state characterised by the inter-particle separation $X$, the equilibrium propagator (giving the probability of a centre-of-mass displacement $x$ in time $\tau$) is
\begin{eqnarray}
\label{greens1}
	G_{X}(x,\tau)&=& A\,\sum_{\alpha} \int_{0}^{\tau} \,p_{\alpha}(\tau-t)\,
	G_{\alpha}(x-x_{\alpha}, t)\, \td t .
\end{eqnarray} 
The summation is over all allowed transitions from the given initial state and $x_{\alpha}$ is the net flux acquired by the transition to state $\alpha$. The coefficient  
$A= [1-e^{-4\omega \tau}]^{-1}$ ensures that the propagator is properly normalised. As in \cite{JPhysA} we define, for each state $i$, the quantity 
\begin{equation}
	m_{i}(\nu,\tau) \equiv \ln \int p_{\tau}^{\rm eq}(J|i)\,e^{\nu \tau J} dJ
\end{equation}
(from which $q_X(\nu)$ will be found). In the present lattice-based model, the net flux assumes discrete (half-integer) values, hence,
\begin{equation}
\label{defm}
	m_{i}(\nu,\tau) = \ln \sum_{x=-\infty}^{\infty} G_{i}(x,\tau)\, e^{\nu x}
\mbox{.}
\end{equation} 
From eqn.(\ref{greens1}) we have
\begin{eqnarray*}
\label{eofqs}
	e^{m_{X}(\nu, \tau)} \equiv \sum_{x} G_{X}(x,\tau) e^{\nu x}
	&=& A\, \sum_{x}\sum_{\alpha} \int_{0}^{\tau} \,p_{\alpha}(\tau-t)\,
	G_{\alpha}(x-x_{\alpha}, t)\,e^{\nu x}\, \td t 	\nonumber \\
	&=&  A\,\sum_{\alpha} \int_{0}^{\tau}\,\omega\,e^{4\omega(t-\tau)}\,
	e^{m_{\alpha}(\nu,t)}\,e^{\nu x_{\alpha}}\, \td t
\end{eqnarray*}
In the steady-state limit $\tau \rightarrow \infty$, we have $A\to1$,
$\partial m_{\beta}(\nu,\tau)/\partial\tau \to Q(\nu)$ \cite{JPhysA}, and  $m_{i}(\nu,\tau)-m_{j}(\nu,\tau) \to q_{i}(\nu)-q_{j}(\nu)$ (see eqn.(\ref{can2})), so differentiating the above equation gives
\begin{eqnarray}
	e^{q_{X}(\nu)}\, (Q+4\omega)&=& \omega \sum_{\alpha} 
	e^{q_{\alpha}(\nu)}\, e^{\nu x_{\alpha}}\nonumber \\
	&=& \omega \left[ e^{q_{X+1}}\, e^{\,\nu/2}+e^{q_{X-1}}\, e^{\,\nu/2}
	+ e^{q_{X+1}}\,e^{-\,\nu/2}+e^{q_{X-1}}\,e^{-\,\nu/2}\right] 	\nonumber \\
	&=& 2\omega \cosh(\nu/2)\left[e^{q_{X+1}}+e^{q_{X-1}}\right]
\label{recurrence1}
\end{eqnarray}
For the special case $X=1$, only transitions to states with $X=2$ are possible, so 
\begin{equation}
\label{recurrence2}
	e^{q_{1}(\nu)}\, (Q+2\omega) = 2\omega \cosh(\nu/2)\, e^{q_{2}(\nu)} \; .
\end{equation}
As $X \rightarrow \infty$, $q_{X}(\nu)$ becomes independent of $X$. Using this in eqn.(\ref{eofqs}) yields

\begin{equation*}
Q + 4\omega= 2\omega \cosh(\nu/2) \times 2 
\end{equation*}
\begin{equation}
 \Longrightarrow Q =4\omega\left[\cosh(\nu/2)-1 \right]
\end{equation}
Equations (\ref{recurrence1}), (\ref{recurrence2}) along with the
above expression for $Q$ are sufficient to determine the quantities 
$q_{i}(\nu)-q_{j}(\nu)$ required in Eqs.~(\ref{omega1R}-\ref{omega2L}). Specifically, for states $i=X+1$ and $j=X$ they yield 
\begin{equation}
\frac{e^{q_{X+1}(\nu)}}{e^{q_{X}(\nu)}}= \frac{(X+1) \cosh (\nu/2)- X}{X\cosh(\nu/2)-(X-1)} \mbox{  ,}
\end{equation}
\begin{figure}
  \begin{center}
  	\vspace{-8mm}
    \epsfig{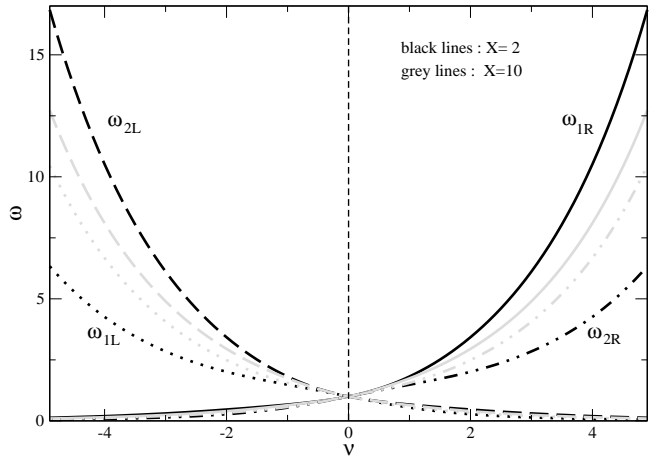}
    \caption{\label{rateb}Transition rates as a function of $\nu$ for $X=2$, $X=10$.
    		Solid lines: $\omega_{1R}$; dashed lines: $\omega_{2L}$; dash-dotted lines:
    		$\omega_{2R}$; dotted lines: $\omega_{1L}$.}
  \end{center}
\end{figure}
and hence the following rates for the four transitions from a state with separation $X$: 
\begin{eqnarray}
	\omega_{1R} &=& \omega e^{\nu/2}\,\frac{(X+1)
	 \cosh(\nu/2)- X}{X\cosh(\nu/2)-(X-1)},		\nonumber \\
	\omega_{1L} &=& \omega e^{-\nu/2}\,\frac{(X-1) 
	 \cosh(\nu/2)-(X-2)}{X\cosh(\nu/2)-(X-1)}\,;  0 \mbox{  for  } 
        X = 1,	\nonumber \\
	\omega_{2R} &=& \omega e^{\nu/2}\,\frac{(X-1)
	 \cosh (\nu/2)- (X-2)}{X\cosh(\nu/2)-(X-1)} \,;  0 \mbox{  for  } 
        X = 1,		\nonumber \\
	 \omega_{2L} &=& \omega e^{-\nu/2}\,\frac{(X+1)
	  \cosh(\nu/2)- X}{X\cosh(\nu/2)-(X-1)}.
\end{eqnarray}

\subsection{Features of the model}

\begin{figure}
  \begin{center}
    \vspace{-10mm}
    \epsfig{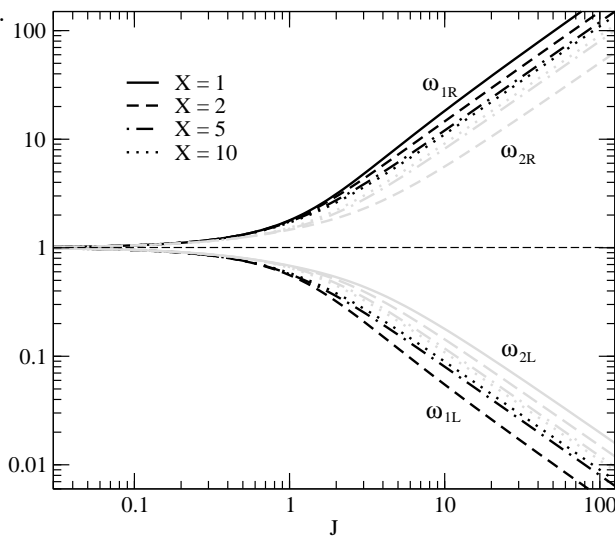}
    \caption{\label{wvsj3}Rates as a function of flux J, on a log scale. The lines
    in grey are for $\omega_{2R}$, $\omega_{2L}$ as shown.}
  \end{center}
\end{figure}

Plotted in Figure \ref{rateb} are the above rates for separations 
$ X= 2,\,10$ as a function of $\nu$. At $\nu =0$ the system is in equilibrium, and all four rates are equal. On increasing the driving 
(acting to the right), hops to the right become more frequent while 
those to the left become rarer. The rates have the symmetries 
$\omega_{1R}(\nu) = \omega_{2L}(-\nu)$ and $\omega_{2R}(\nu) = \omega_{1L}(-\nu)$.
The same rates are plotted as a function of $J$ on a log-log plot (Figure ~\ref{wvsj3}). All curves are linear for large $J$ with positive slopes for hops to the right and negative ones for those to the left. It follows that, in this regime the rates show a power law dependence on the current J, becoming proportional to it  as $X \rightarrow \infty $ for hops to the right and inversely proportional to it for hops to the left. 

Notice that, in both Figs.~\ref{rateb} and \ref{wvsj3}, the rates of hops to the right vary considerably with the separation $X$, being greater for small $X$ for the particle ahead ($1$) and greater for large $X$ for the particle behind ($2$).
For positive $\nu$, particle $2$, in effect, pushes particle $1$ ahead, thereby making hops 
to the right more frequent for it. Particle $1$ in turn pushes $2$ in 
the opposite direction hence suppressing its hop to the right. The interaction decreases as the separation $X$ increases and the rates $\omega_{1R} \rightarrow \omega_{2R} $ as $X \rightarrow \infty$.  Since the {\em bare} interaction exists only when the particles are on adjacent sites, this dependence reveals the existence of a repulsive force of longer range, mediated by the reservoir, 
i.e.~the coloured noise from the non-equilibrium reservoir tends to push particles apart, so that encounters in which the rear particle is prevented from moving forwards, by the direct exclusion interaction, are rarer than from equilibrium noise. 

For the particle-hopping model, this non-equilibrium entropic repulsion is somewhat analogous to the Cassimir effect. 
The force derives from the fact that proximity (and therefore increased likelihood of collision) of a particular pair of particles restricts the fluctuations available to the other, weakly-coupled systems in the ensemble, which are required to meet the flux constraint on average. 
However, in the context of a model of a complex fluid under shear (a more physically-motivated application of NCDB), a reservoir-mediated long-range effective interaction of this kind would be interpreted as a stress field; certain particle motions are encouraged by the configuration of other distance particles, because shear stresses propagate through a system on which a shear flux is imposed. Although such a stress field is not a relevant order parameter of the equilibrium fluid, it would be automatically generated by the present formalism when the equilibrium description is converted into a driven model, because stress-induced transitions are crucial to achieving the required strain rate.
 
A detailed description of non-equilibrium-reservoir-mediated interaction forces is presented in the next section, where we shall see that the symmetries respected by the NCDB formalism yield effective interaction forces that respect Newton's third law of motion.

\section{Effective interaction force between particles in the driven steady state}
\label{Newton}

Let us now use the intuition gained from the two-particle exclusion model (section \ref{exclusion}) to consider the effective interactions that arise when the NCDB is applied to a more general system with a particle current.
In a driven steady state like that modelled in section \ref{exclusion}, particles are subject to forces arising from interactions with one another and from the reservoir that maintains the system in the driven steady state. Their motion is stochastic and the evolution of their distribution is given (for a continuum-space model) by the Fokker Planck equation with drift and diffusion co-efficients $A(x)$ and $D$ respectively:
\begin{equation}
\label{fpe1}
	\partial_t P(x,t)= \partial_x\left(-A(x) + D\,\partial_x \right) P(x,t)
\end{equation}
Here $P(x,t)$ is the probability that a particle is at position $x$ at time $t$ subject to its position being $x_0$ at an earlier time $t_0$. For short times, the probability for a particle to drift from its position $x_0$ to $x$ in time $\Delta$ can be expressed in terms of $A(x)$ and $D$ as
\begin{equation}
\label{trans1}
	P(x,t+ \Delta|\,x_0, t)= \frac{1}{2\sqrt{\pi D \Delta}}
	\exp\left(-\frac{(x-x_0-A(x_0) \Delta)^2}{4D \Delta}\right) \; .
\end{equation}
We identify the rate for this {\it transition} (in the language of {\em discretized} states) to be
\begin{equation}
	\omega_{x_0 \rightarrow x} =  \frac{P(x,t+ \Delta|\,x_0, t)}{\Delta\,}.
\end{equation}
In two or more dimensions, with isotropic mobility, eqn.(\ref{fpe1}) generalises to
\begin{equation}
	\partial_t P({\bf x},t)= -{\bf\nabla} \cdot\left({\bf A(x)} 
	+ D\,{\bf\nabla}\right) P({\bf x},t) .
\end{equation}

A Fokker-Planck equation of this kind might be derived from the statistics of a particular stochastic system (for instance, a system whose stochastic rates are determined by NCDB). Whatever its origin, when we encounter a Fokker-Planck equation of this form, we can interpret the drift vector ${\bf A(x)}$ as being proportional to an effective force on the particle, ${\bf F(x)}\equiv k{\bf A(x)}$ (where the constant of proportionality $k$ is some friction coefficient). Correspondingly the transition probability is 
\begin{equation}
	P({\bf x},t+ \Delta|\,{\bf x_0}, t)= \frac{1}{(2\sqrt{\pi D \Delta})^3}
	\exp-\left(\frac{({\bf  x}- {\bf x_{0}}-{\bf A(x_0)} \Delta)^2}{4D \Delta}
         \right) \mbox{.}
\end{equation}
It follows, for transitions 
$\bf{x_0} \rightarrow \bf{x} = \bf{x_0 } + \bf{\delta x}$ and 
$\bf{x_0} \rightarrow x^\prime = \bf{x_0 } - \bf{\delta x}$, that the ratio of the rates is a simple function of ${\bf A}({\bf x_0})$ and the (small) displacement $\bf{\delta x}$,
\begin{equation}
\label{ratio}
  \frac{\omega_{\bf{x_0} \rightarrow \bf{x}}}
  {\omega_{\bf{x_0}\rightarrow \bf{x}^{\prime}}} 
  =  \exp \left(\frac{{\bf A(x_0)}\cdot{\bf \delta x}}{D}\right) .
\end{equation}
This expression holds whether the reservoir and other particles (giving rise to the forces ${\bf F(x)}$) are in an equilibrium or non-equilibrium steady-state.
Given the rates for these transitions in the driven state (eqn.~(\ref{can1})), we can calculate the effective force of interaction between particles.  

Consider two particles, possibly surrounded by many others that comprise the system, which interact with each other, both directly, and via the rest of the system, and  via the reservoir (in the driven case). Particles 1 and 2 are located at ${\bf x}_1$ and ${\bf x}_2$ respectively, and the {\em total} net force on either can be described by $\bf A(x)$ in the Fokker-Planck equation.
\begin{figure}
\label{sym}
\begin{center}
   \epsfig{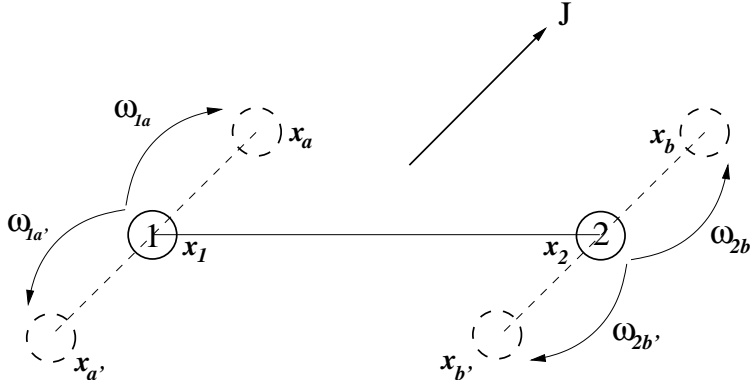}
    \caption{Transitions $1a\,$,$\,1a^{\prime}$ and $2b\,$,$\,2b^{\prime}$ of
     particles $1$ and $2$ respectively.}
\end{center}
\end{figure}

For a transition whereby particle 1 (as shown in fig.~\ref{sym}.) moves from its initial position to positions ${\bf x}_a$ or ${\bf x}_{a\prime}$ along the direction of the average flux ${\bf J}$, the ratio of the rates implied by eqns.~(\ref{ratio}) and (\ref{can1}) is given by
\begin{equation}
\label{logratio}
	\ln \left(\frac{\omega_{1a}}{\omega_{1a^\prime}}\right)=
	\ln\left(\frac{\omega^{\rm eq}_{1a}}{\omega^{\rm eq}_{1a^\prime}}\right) 
	+\nu \delta + q_{1a}-q_{1a^\prime} = \frac{A^{\parallel}({\bf x}_1) \delta}{D} \; ,
\end{equation}
yielding $A^{\parallel}(\bf{x})$, the component of $\bf{A(x)}$ parallel to $\bf{J}$. Here, the centre-of-mass displacement due to the transition ${\bf x}_1 \rightarrow {\bf x}_a$ is $\delta= |\delta {\bf x}|$ and the states $1a$ and $1a^\prime$ correspond to particle 1 being at ${\bf x}_a$ or ${\bf x}_{a^\prime}$ respectively, while particle 2 remains at ${\bf x}_2$. Similarly, for particle 2,
\begin{equation}
	\ln\left(\frac{\omega_{2b}}{\omega_{2b^\prime}}\right)=
	\ln\left(\frac{\omega^{\rm eq}_{2b}}{\omega^{\rm eq}_{2b^\prime}}\right) 
	+ \nu \delta + q_{2b}-q_{2b^\prime} = \frac{A^{\parallel}({\bf x}_2) \delta}{D}
\end{equation}
where now the states $2b$ and $2b^\prime$ correspond to particle 2 being at ${\bf x}_b$ and ${\bf x}_{b^\prime}$ respectively while particle 1 remains at its initial position ${\bf x}_1$. 

Let us now invoke translational invariance. If these two particles interact with each other via a potential that depends only on their relative separation, and their coupling to the non-equilibrium reservoir, which drives them, is also independent of their absolute position, then state $1a$ becomes equivalent to $ 2b^\prime$ and similarly $1a^\prime \equiv 2b$. (Note that we do not assume any rotational symmetry, since it is broken by the anisotropic flux.) We see that translational invariance, in our reservoir-driven two-particle systems, leads to a total effective force acting on particle 1 with component $F^{\parallel}= k A^{\parallel}$ in the direction of the flux $\bf{J}$ and is given by
\begin{eqnarray}
	F^{\parallel}_{1}&=& k A^{\parallel}_{1}=k(q_{1a}- q_{1a^\prime}+\nu \delta)=
	k[-(q_{2b}-q_{2b^\prime})+\nu \delta]\\
	&=&-(F^{\parallel}_{2} - k\nu \delta) +k\nu \delta \mbox{ .}  
\label{fint}
\end{eqnarray}
The contribution $\nu \delta$ to the effective force does not
depend on the presence of the other particle and is therefore a steady, monopole force on each from the reservoir. Defining the force of interaction along the direction of $\bf{J}$ as
\begin{equation}
	F^{\parallel}_{\rm int}(x) \equiv F^{\parallel}(x) - k\nu \delta ,
\end{equation}
we find that equation (\ref{fint}) implies
\begin{equation}
\label{N3A}
	F^{\parallel}_{\rm int 1}=- F^{\parallel}_{\rm int 2} \mbox{ .}
\end{equation}
Following the same procedure, it can be shown that eqn.(\ref{fint}) also holds for hops perpendicular to ${\bf J}$, i.e.,
\begin{equation}
\label{Fperp}
	F_{1}^{\perp}= k A_{1}^{\perp}=k(q_{1a}- q_{1a^\prime}+0)=
	k[-(q_{2b}-q_{2b^\prime})+0]
\end{equation}
These transitions do not carry any flux along ${\bf J}$ (the terms corresponding to $2k\nu \delta$ in the earlier expressions vanish). It follows from Eq.~(\ref{Fperp}) that 
\begin{equation}
\label{N3B}
	F_{1}^{\perp}=-  F_{2}^{\perp} \mbox{.}
\end{equation}
We see that the effective forces of interaction between the two particles, which are mediated by the non-equilibrium ensemble, respect Newton's third law of motion. This results from the translational invariance of the system, and from the microscopic reversibility which is built into the NCDB formalism, despite being broken by the imposed macroscopic flux constraint.

\subsection{Effective forces in the two-particle exclusion model}

\begin{figure}	
	\begin{center}
  	\vspace{-8mm}
		\epsfig{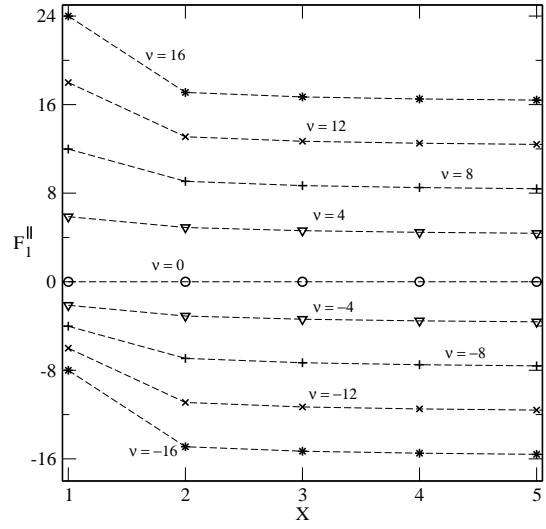}
		\caption{\label{force}The effective force on particle 1, 
		$F_1^{\parallel}(\nu, X)$ for some values of $\nu$.}
	\end{center}
\end{figure}

\begin{figure}
	\begin{center}
  	\vspace{-8mm}
  	\epsfig{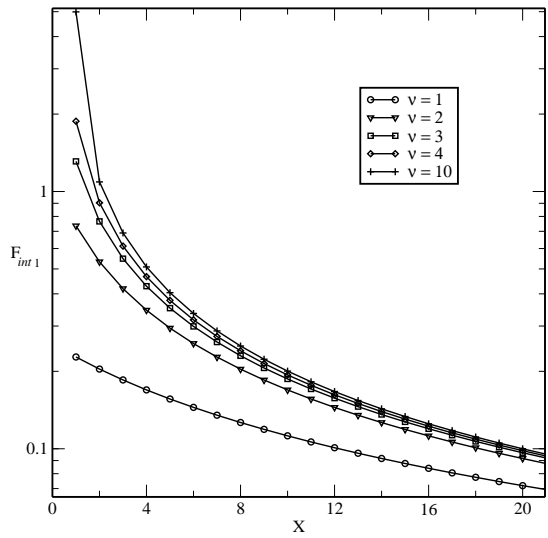}
		\caption{\label{scal}Effective interaction force $F_{\rm int\,1}(\nu, X)$
		for some values of $\nu$.}
  \end{center}
\end{figure}

Applying the above discussion to our results from the two-particle exclusion model (section \ref{exclusion}) allows us to convert the hopping rates, that NCDB dictates for that model, into effective reservoir-mediated interaction forces. We find
\begin{eqnarray}
	F_{1}^{\parallel}(\nu, X) 
	&\propto& \ln \left(\frac{\omega_{1R}}{\omega_{1L}}\right)
	=  \nu +\ln \left[\frac{(X+1) \cosh (\nu/2)-X}
	{(X-1)\cosh(\nu/2)-(X-2)}\right] \mbox{  ,}	\nonumber \\
	F_{2}^{\parallel}(\nu, X) 
	&\propto& \ln \left(\frac{\omega_{2R}}{\omega_{2L}}\right)
	= \nu +\ln \left[\frac{(X-1) \cosh (\nu/2)-(X-2)} {(X+1)\cosh(\nu/2)-X}\right] \mbox{  .}
	\label{Feff}
\end{eqnarray}
The effective force on both particles tends to a constant value
proportional to $\nu$ for large $X$. This is the monopole contribution from the 
driven reservoir and is equal on both particles. We subtract this contribution from the force on each particle to find the force of interaction. Equations (\ref{Feff}) thus imply
\begin{equation}
  F_{\rm int\,1}^{\parallel}(\nu,X)
  =F_{1}^{\parallel}(\nu, X)-k(\nu)
  =-\left[F_{2}^{\parallel}(\nu, X)-k(\nu)\right]
  = -F_{\rm int\,2}^{\parallel}(\nu,X) 
\end{equation}
as for the general continuum case derived above. Furthermore, from Eq.~(\ref{Feff}), we see that $F_{\rm int\,1}^{\parallel}$ is even in $\nu$ so that changing the direction of driving does not alter the two-particle effective interactions. The effective force $F_{1}^{\parallel}(\nu, X)$ and the interaction force $F_{\rm int\,1}^{\parallel}(\nu,X)$ on particle 1 are plotted for some values of $\nu$ in Figures \ref{force} and 
\ref{scal} respectively. 

We notice an additional invariance of this interaction force, peculiar to the two-particle exclusion model. We find that $F_{\rm int\,1}^{\parallel}(\nu,X+M)$ for any value of $\nu$ is equal to $F_{\rm int\,1}^{\parallel}(1,X)$ for
a specific value $M$ of translation. So, increasing the driving force $\nu$ increases the range of the effective interaction, in such a way that the tail of the repulsive force is shifted, but not altered in shape. To see this from the expressions above:
\begin{eqnarray}
	F_{\rm int\,1}^{\parallel}(1,X)=\ln \left[\frac{(X+1) \cosh (1/2)-X}
	{(X-1)\cosh(1/2)-(X-2)}\right] {\mbox ,}\\
	F_{\rm int\,1}^{\parallel}(\nu,X+M)=\ln \left[\frac{(X+1+M)\cosh (\nu/2)-(X+M)}
	{(X-1+M)\cosh(\nu/2)-(X+M-2)}\right] {\mbox .}
\end{eqnarray}
Equating the two,
\begin{equation}
	\left[\frac{(X+1) \cosh (1/2)-X}{(X-1)\cosh(1/2)-(X-2)}\right]
	= \left[\frac{(X+1+M)\cosh (\nu/2)-(X+M)}{(X-1+M)\cosh(\nu/2)-(X+M-2)}\right]
\end{equation}
yields the following expression for the translation $M$ as a function
of $\nu$:
\begin{equation}
  M = \frac{\cosh(\nu/2)-\cosh(1/2)}
  {\left[1-\cosh(\nu/2)\right]\left[1-\cosh(1/2)\right]}
\end{equation}
The curves $F_{\rm int\,1}^{\parallel}(\nu,X)$ for $ \nu = 2,3,4,10$ shown in Figure \ref{scal} collapse onto $F_{\rm int\,1}^{\parallel}(1,X)$ for translation  $M= 5.994,\,7.096,\,7.473,\,7.6405,\,7.7251$ respectively. The reason for this particular invariance is a mystery to us.

\section{Discussion and Conclusions}

The derivation of the non-equilibrium dynamics in section \ref{foundations} (denoted NCDB \cite{PRL,JPhysA,PhysicaA}) relies only on the existence of a steady state, mechanically boundary-driven by a weakly-coupled reservoir in a microscopically reversible system (the prime example being a complex fluid under shear),  with no other assumptions or approximations. We have argued here that such driven steady states do exist, since many complex fluids are found in non-equilibrium states where the bulk properties are insensitive to boundary conditions and measuring times, so that an explicit treatment of dissipation can legitimately be neglected (despite its importance in other types of steady states \cite{Blythe07}). Thus, the theory does not apply to systems in which the dynamics at the microscopic level is affected by the rate of heating caused by the driving force and excludes systems driven by a field in the bulk (e.g.~electrophoresis) or by a temperature gradient, or to systems with irreversible microscopic dynamics such as granular media and molecular motors. The remaining class to which the theory applies is nonetheless large, including most complex fluids under shear, which exhibit a huge variety of non-equilibrium structures and transitions. One such example is discussed in section \ref{intro}; another is a sheared foam, for which an effective temperature much higher than ambient is required \cite{Sollich98,Evans99} to theoretically model the extra noise generated by the foam itself when it flows. That a firm statistical foundation now exists for the study of such non-equilibrium systems is a significant advance.

The formalism that results from the foundations in section \ref{foundations} was previously derived from a Bayesian interpretation of probabilities (``MaxEnt") in Ref.~\cite{JPhysA} (that interpretation also appearing in Ref.~\cite{PRL}). While a firmer foundation for the physics has now been provided, the resulting mathematical structure \cite{JPhysA} remains unchanged, and we have applied it, in section \ref{exclusion}, to a model of symmetric exclusion on a one dimensional lattice. We have found the rates for the various transitions in terms of a single parameter $\nu$ determining the average flux in the driven state. These rates allow us to quantify an effective interaction force between two particles in the driven state, mediated by the non-equilibrium reservoir. Due to the symmetries inherent in the NCDB formalism, this force obeys Newtons third law of motion. We shall demonstrate the experimental validity of NCDB in a future publication.

\medskip
\noindent{\bf Acknowledgements:} We are grateful to Peter Olmsted and Richard Blythe for enlightening discussions. The work was funded by EPSRC grant GR/T24593/01. RMLE is supported by the Royal Society.

\section*{Appendix}

We derive here the expression for the driven transition rates Eq.~(\ref{can1}) from the probability of a trajectory $\Gamma$ in the driven ensemble (Eq.~(\ref{proportional})): $p^{\rm driven}(\Gamma)\propto p^{\rm equilib}(\Gamma)e^{\nu\gamma(\Gamma)}$.
The transition rate between two microstates $a$ and $b$ is defined in terms of the probability $\Pr(b|a)$ of making the transition to state $b$ within a small time interval $\Delta t$, given that the system is in state $a$: $\omega_{a\rightarrow b}=\Pr(b|a)/\Delta t$. (This is most easily understood as the continuum limit of a discrete-time process, with time step $\Delta t$.) In turn, this transition probability is obtained by summing over those trajectories $\Gamma$ that contain the states $a$ and $b$ consecutively, weighted by $p(\Gamma)$ 
\begin{eqnarray}
  Pr(b|a)  = \frac{\Pr(a,b)}{\Pr(a)}
  = \frac{\sum_{\Gamma\ni(a,b)}p(\Gamma)}{\sum_{\Gamma\ni a}p(\Gamma)}.
\end{eqnarray}
Here, $\Pr(a,b)$ is the combined probability of finding the system in state $a$ at a given time (defined, without loss of generality, to be $t=0$), and in state $b$ on the subsequent time step $t=\Delta t$, while $\Pr(a)$ is simply the microstate occupancy. The sum $\sum_{\Gamma\ni a}$ denotes a summation over all trajectories that pass through microstate $a$ at $t=0$. Likewise $\sum_{\Gamma\ni(a,b)}$ sums trajectories containing state $a$ at $t=0$ and $b$ at $t=\Delta t$. Thus, using the central result, Eq.~(\ref{proportional}), a transition rate in the driven steady state can be written as
\begin{eqnarray}
\label{A_delta}
  \omega^{\rm dr}_{ab} 
  = \frac{\sum_{\Gamma\ni(a,b)}p^{\rm eq}(\Gamma)e^{\nu\gamma(\Gamma)}} 
  {\Delta t\sum_{\Gamma\ni a}p^{\rm eq}(\Gamma) e^{\nu\gamma(\Gamma)}}
  = \frac{\int_{-\infty}^\infty e^{\nu\gamma}\sum_{\Gamma\ni(a,b)}
  \delta(\gamma,\gamma(\Gamma))\,p^{\rm eq}(\Gamma)\,d\gamma}
  {\Delta t \int_{-\infty}^\infty e^{\nu\gamma}\sum_{\Gamma\ni a}
  \delta(\gamma,\gamma(\Gamma))\,p^{\rm eq}(\Gamma)\,d\gamma},
\end{eqnarray}
where $\gamma(\Gamma)$ is the total shear-strain undergone by a system following trajectory $\Gamma$. In the numerator, the average over the Dirac-delta $\delta(\gamma,\gamma(\Gamma))$ yields the probability distribution for total shear-strain, $p_\tau^{\rm eq}(\gamma|a,b)$, conditioned on the transition occurring,
\begin{eqnarray}
  \sum_{\Gamma\ni(a,b)}\delta(\gamma,\gamma(\Gamma))\,p^{\rm eq}(\Gamma) = 
  p_\tau^{\rm eq}(\gamma|a,b)\sum_{\Gamma\ni(a,b)}p^{\rm eq}(\Gamma).
\end{eqnarray}
Similarly, the denominator in Eq.~(\ref{A_delta}) leads to an expression containing $p_\tau^{\rm eq}(\gamma|a)$. The subscript $\tau$ denotes the implicit dependence of these distributions on the duration of the trajectory $\Gamma$. After factoring out the equilibrium rate $\omega^{\rm eq}_{ab}=\sum_{\Gamma\ni(a,b)}p^{\rm eq}(\Gamma)/\sum_{\Gamma\ni a}p^{\rm eq}(\Gamma)$, we can write the driven transition rate as
\begin{eqnarray}
\label{A_drivenDB}
  \omega^{\rm dr}_{ab} = \omega^{\rm eq}_{ab}\;\lim_{\tau\rightarrow\infty}
  \frac{\int_{-\infty}^\infty p_\tau^{\rm eq}(\gamma|a,b)\,e^{\nu\gamma}\,d\gamma}
  {\int_{-\infty}^\infty p_\tau^{\rm eq}(\gamma|a)\,e^{\nu\gamma}\,d\gamma}.
\end{eqnarray}
Here, $p_\tau^{\rm eq}(\gamma|a,b)$ is the distribution for the probability that the system at equilibrium accumulates a total amount of shear $\gamma$ over a time period $\tau$ due to equilibrium fluctuations, given that it made a transition from $a$ to $b$. The $\tau\rightarrow\infty$ limit guarantees that the system has attained its stationary state, where total shear and average shear flux $J$ are related by $\gamma=J\tau$. 

Equation~(\ref{A_drivenDB}) is the \textit{canonical-flux} representation of a transition rate $a\rightarrow b$ in the driven ensemble, which was derived from an information-theoretic principal of least constraint in \cite{JPhysA}. The $\tau$-independent representation (Eq.~(\ref{can1})) follows from this as shown in \cite{JPhysA}.

\end{document}